\documentclass[aps,prb,amssymb,twocolumn,showpacs]{revtex4}
\usepackage{graphicx}

\newcommand{\bea}{\begin{eqnarray}}

\newcommand{\eea}{\end{eqnarray}}

\newcommand{\beq}{\begin{equation}}

\newcommand{\eeq}{\end{equation}}

\newlength{\textwidthm}

\begin{document}

\def \tr{{\mbox{tr~}}}
\def \ra{{\rightarrow}}
\def \ua{{\uparrow}}
\def \da{{\downarrow}}
\def \be{\begin{equation}}
\def \ee{\end{equation}}
\def \ba{\begin{array}}
\def \ea{\end{array}}
\def \bea{\begin{eqnarray}}
\def \eea{\end{eqnarray}}
\def \nn{\nonumber}
\def \l{\left}
\def \r{\right}
\def \half{{1\over 2}}
\def \etal{{\it {et al}}}
\def \cH{{\cal{H}}}
\def \cM{{\cal{M}}}
\def \cN{{\cal{N}}}
\def \cQ{{\cal Q}}
\def \cI{{\cal I}}
\def \cV{{\cal V}}
\def \cG{{\cal G}}
\def \cF{{\cal F}}
\def \cZ{{\cal Z}}
\def \bS{{\bf S}}
\def \bI{{\bf I}}
\def \bL{{\bf L}}
\def \bG{{\bf G}}
\def \bQ{{\bf Q}}
\def \bK{{\bf K}}
\def \bR{{\bf R}}
\def \br{{\bf r}}
\def \bu{{\bf u}}
\def \bq{{\bf q}}
\def \bk{{\bf k}}
\def \bz{{\bf z}}
\def \bx{{\bf x}}
\def \bpsi{{\bar{\psi}}}
\def \tJ{{\tilde{J}}}
\def \W{{\Omega}}
\def \e{{\epsilon}}
\def \lam{{\lambda}}
\def \L{{\Lambda}}
\def \a{{\alpha}}
\def \t{{\theta}}
\def \b{{\beta}}
\def \g{{\gamma}}
\def \D{{\Delta}}
\def \d{{\delta}}
\def \w{{\omega}}
\def \s{{\sigma}}
\def \f{{\varphi}}
\def \x{{\chi}}
\def \e{{\epsilon}}
\def \h{{\eta}}
\def \G{{\Gamma}}
\def \z{{\zeta}}
\def \hatt{{\hat{\t}}}
\def \hn{{\bar{n}}}
\def \vk{{{\bf k}}}
\def \vq{{{\bf q}}}
\def \gk{{\g_{\vk}}}
\def \nd{{^{\vphantom{\dagger}}}}
\def \yd{^\dagger}
\def \av#1{{\langle#1\rangle}}
\def \ket#1{{\,|\,#1\,\rangle\,}}
\def \bra#1{{\,\langle\,#1\,|\,}}
\def \braket#1#2{{\,\langle\,#1\,|\,#2\,\rangle\,}}

\title{Exotic Superconducting Phases of Ultracold Atom Mixtures on Triangular
 Lattices}

\author{L.~Mathey$^1$, S.~-W.~Tsai$^2$, and A.~H.~Castro~Neto$^3$}

\affiliation{$^1$Physics Department, Harvard University, Cambridge, MA 02138 \\
$^2$Department of Physics and Astronomy, University of California, Riverside, CA 92521 \\
$^3$Department of Physics, Boston University, 590 Commonwealth Ave., Boston, MA 02215
}

\date{\today}

\begin{abstract}
We study the phase diagram of two-dimensional Bose-Fermi mixtures of
ultracold atoms on a triangular optical lattice, in the limit 
when the velocity of bosonic condensate fluctuations is much larger 
than the Fermi velocity. 
 We contrast
 this work with our previous results 
 for a square lattice system  in 
  Phys. Rev. Lett.  {\bf 97}, 030601 (2006).
 Using functional renormalization group techniques
we show that the phase diagrams for a triangular
 lattice  contain 
exotic superconducting phases. For spin-$1/2$ fermions on an
isotropic lattice we find a competition of $s$-, $p$-, extended
$d$-, and $f$-wave symmetry, as well as antiferromagnetic order.
For an anisotropic lattice, we further find an extended $p$-wave phase.
A Bose-Fermi mixture with spinless  fermions on an isotropic lattice shows 
a competition between $p$- and $f$-wave symmetry. 
 These phases can be traced back to the geometric shapes
 of the Fermi surfaces in various regimes, as well as the intrinsic 
 frustration of a triangular lattice. 
\end{abstract}


\maketitle

\section{Introduction}\label{intro}
During the last decade there has been remarkable progress in controlling and 
manipulating ensembles of ultracold atoms \cite{stoeferle, mandel, koehl}.
Using optical lattices and different types of atoms,
a variety of non-trivial many-body states can be 'engineered' that are very accurate
realizations of Hubbard models. The experimental observation of 
the Mott-Hubbard superfluid-insulator transition \cite{greiner},
of fermionic superfluids \cite{BECBCS}, 
of the Tonks-Girardeau gas \cite{Tonks} and other
Luttinger liquids \cite{Luttinger}, the 
measurement of noise correlations \cite{noise}, 
instabilities \cite{inst}  as well as localization effects \cite{BFMlocal} in Bose-Fermi mixtures (BFM),
have created a great deal of excitement in the physics community. 
These findings have prompted a large amount of theoretical work
that study charge density wave (CDW) order \cite{cdw}, 
many-body states of composite particles \cite{composite}, polaronic effects
\cite{mathey}, and tuning of Feshbach resonances on lattices \cite{ho}.

It is known that the geometric shape of the lattice is a crucial factor
in determining the properties of interacting many-body systems. For instance,
localized spins interacting antiferromagnetically on a triangular lattice
suffer from the phenomenon of {\it frustration}, when antiferromagnetic order cannot
be achieved because of the particular lattice structure. 
For itinerant fermionic systems, the lattice structure, together with the 
dispersion relation and the filling fraction, determine the shape of the
Fermi surface (FS). 
The FS, by its turn, is a crucial factor in determining what type of orders the
system can develop. Indeed, for the triangular lattice we consider in this
paper, which shows a rich and subtle competition between superconducting phases
with different symmetries, we find
that small changes in the shape of the FS determine which pairing symmetry
is dominant. This is a reflection of the ``lattice frustration'' on the
superconducting phases.
In solids, this intriguing lattice geometry is realized in materials such as cobaltates
\cite{cobaltates}, transition metal dichalcogenides \cite{dichal} and $\kappa$-(ET)$_2$X
layered organic crystals  \cite{jerome} (if each lattice site is represented by one ET dimer
\cite{kino}), and has been the subject of several theoretical studies
\cite{tri_weak,tri_strong}.

The versatile technology of optical lattices allows to create and tune a vast
number of lattice geometries for ultracold atoms, and is therefore a
promising experimental environment to study the rich many-body physics that
can be engineered within these systems.  We study mixtures of ultracold fermionic atoms
(either with two hyperfine states that are labeled by pseudo-spin $\ua$ and
$\da$, or single component, spinless, fermions), and single component bosons
on a  triangular lattice. Such a system can be realized as a
$^{40}$K-$^{87}$Rb mixture 
in an optical lattice created by Nd:YAG lasers, with tunable interactions.
The geometry of the lattice under consideration 
 is shown in Fig. \ref{phases} a).

This paper
 is organized as follows:
 In section \ref{effHam} we derive the effective Hamiltonian for 
 the fermionic degrees of freedom, in the limit of fast condensate
 modes. In section \ref{FRG}
  we describe the functional renormalization group approach and its
 implementation.
 In section \ref{iso} and \ref{aniso} we describe BFMs with
 spin $1/2$ on an isotropic and an anisotropic lattice, respectively,
and in section \ref{nospin} we consider spinless fermions.
 In the last section \ref{conclude} we comment on the 
 detection of the phases found in this paper, and conclude.

\section{Effective Hamiltonian}\label{effHam}
\begin{figure}
\includegraphics[width=8.0cm]{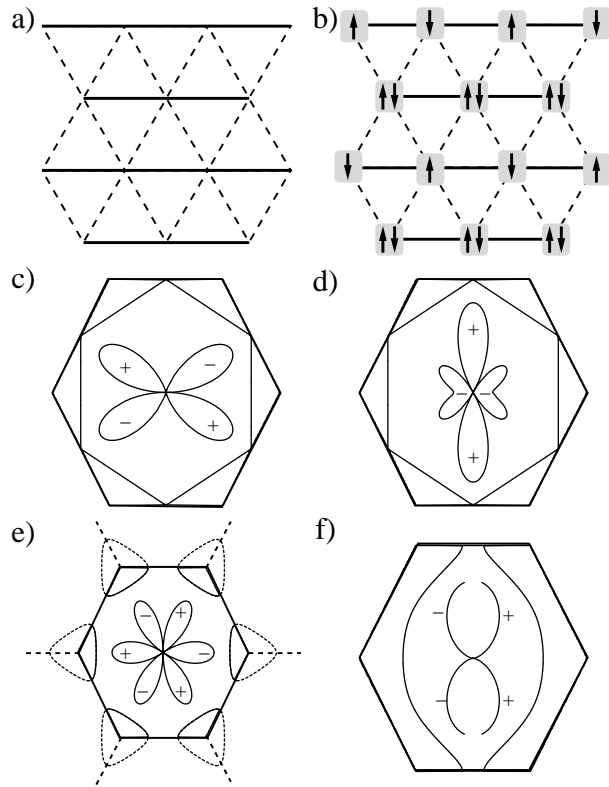}
\caption{\label{phases}
 a) Lattice geometry of the system.  
 The continuous (dashed) bonds correspond to the hopping amplitudes
 $t_{b/f, 1 (2)}$. 
 For $t_{b/f, 1}=t_{b/f, 2}$, the lattice is an isotropic triangular lattice.
  b) schematic representation of the AF order corresponding to nesting
 vector ${\bf Q}_1$.  
 c) + d) Order parameters of  the extended $d$-wave orders $D_1$ and $D_2$.
 e) Order parameter of the $f$-wave phase. This order can also be interpreted
 as two $s$-wave paired hole states whose order parameters are out of phase
 by $\pi$.  f) Order parameter of the extended 
 $p$-wave phase, that appears in anisotropic lattices. 
}
\end{figure}

%
%
%
%
BFMs in optical lattices are well described by
a Hubbard model of the form, 
\bea\label{Ham}
H & = & - \sum_{\langle i j \rangle_{n=1,2},s} t_{f,n} f^\dagger_{i,s} 
f_{j,s} -  \sum_{\langle i j \rangle_{n=1,2}} t_{b,n} b^\dagger_{i} b_{j}\nonumber\\
& & - \sum_i (\mu_{f} n_{f,i} + \mu_b n_{b,i})
+   \sum_{i} \Big[U_{ff} n_{f, i,\uparrow}n_{f,i, \downarrow}\nonumber\\
& & + \frac{U_{bb}}{2} n_{b, i}n_{b,i} + U_{bf} n_{b, i} n_{f,i} \Big] \, ,
\eea
where $f^\dagger_{i,s}$ ($f_{i,s}$) creates (annihilates) a 
fermion at site $i$ with 
pseudo-spin $s$ ($s=\uparrow,\downarrow$), $b^\dagger_i$ ($b_i$) 
creates (annihilates) a boson at site $i$,
$n_{f,i}= \sum_s f^\dagger_{i,s} f_{i,s}$ ($n_{b,i} = b^\dagger_i b_i$) is 
the fermion (boson) number operator.
With this
 Hamiltonian we describe a BFM with spin $1/2$ fermions, whereas 
 a  Hamiltonian of a BFM with spinless fermions can be obtained 
from this Hamiltonian by suppressing the spin indices. 
 Here we allowed for
 two different values for the hopping amplitudes, for
 the two types of lattice bonds $\langle i j\rangle_1$
 and $\langle i j\rangle_2$.
 For a triangular lattice, 
$t_{f,a}$ and $t_{b,a}$ with $a=1,2$ are 
the fermionic and bosonic tunneling amplitudes between neighboring sites, 
where the index $a=1$ ($a=2$) refers to the continuous (dashed) bonds, as
shown in Fig. \ref{phases} a).
 For the description of the isotropic case
 we equate $t_{b/f,1}$ and $t_{b/f,2}$, and
 define $t_f\equiv t_{f,1}=t_{f,2}$ and $t_b\equiv t_{b,1}=t_{b,2}$. 
$\mu_f$ ($\mu_b$) is the chemical potential for fermions (bosons), 
$U_{bb}$, $U_{ff}$, and $U_{bf}$ are the on-site boson-boson, fermion-fermion
and boson-fermion repulsion energy, respectively.

We consider the limit of weakly interacting bosons, in which
 the bosons form a BEC, for which we use  
  the standard Bogoliubov description. 
Specifically, we assume that a macroscopic number $N_0$ of bosons
 is condensed in the $k=0$ mode, which we take into account by
 replacing the operator $b_0$ by $\sqrt{N_0}$.
 We then keep all the remaining terms up to second order in the
 operators $b_\bold{k}$, with $\bold{k}\neq 0$, which includes anomalous terms 
corresponding to pair creation and annihilation from the condensate, and
 diagonalize this expression via a Bogoliubov transformation.
 The resulting  dispersion relation is given by
\bea\label{disp}
 \omega_\bold{k} & = & \sqrt{(\epsilon_{b,\bold{k}} -
 \epsilon_{b,0})(\epsilon_{b,\bold{k}} - \epsilon_{b,0} 
 +2U_{bb}n_b)},
\eea
 where the bare lattice dispersion is given by:
\bea\label{baredisp}
 \epsilon_{b,k} & = &  -t_{b,1} 2 \cos k_x - 
 t_{b,2}(2 \cos(k_x/2 + \sqrt{3}k_y/2)\nonumber\\
& & + 2 \cos(k_x/2 - \sqrt{3}k_y/2)).
\eea 
For small values of $k_x$ and $k_y$, $\omega_{\bold{k}}$ can
 be expanded as:
\bea
 \omega_{\bold{k}} & \sim & \sqrt{((2 t_{b,1}+t_{b,2})k_x^2 + 3 t_{b,2}k_y^2) 
 U_{bb}n_b},
\eea
 which gives us the two velocities
\bea
 v_{b,x}& =&\sqrt{(2 t_{b,1} + t_{b,2})U_{bb}n_b}\\
 v_{b,y} & = &\sqrt{3t_{b,2}U_{bb}n_b}.
\eea
In addition to the assumption that the bosons form a BEC, 
 we assume that these velocities of the 
condensate fluctuations are much larger than the Fermi velocity,  
     which corresponds to the conditions  
 $v_{b,x/y} > t_{f,1/2}$. 
 Therefore, large bosonic hopping amplitudes, 
 a bosonic density of $\approx 1$--$3$, and some intermediate
 value for $t_{f,1/2}$ will satisfy this requirement. 
In this limit the bosonic modes can be integrated out, and one obtains an approximately
non-retarded fermion-fermion interaction.  The induced potential $V_{{\rm
    ind.}, \bold{k}}$ is given by:
\bea
V_{{\rm ind.}, \bold{k}} & = & - \tilde{V}/(1 + \xi_1^2 (2 - 2 \cos k_x)\nonumber\\
 && + 
 \xi_2^2(4 - 4 \cos(k_x/2)\cos(\sqrt{3}k_y/2))) \, ,
\eea
with $\tilde{V}=U_{bf}^2/U_{bb}$, and $\xi_{a}$ are the healing lengths of
the Bose-Einstein condensate (BEC) 
and are given by  $\xi_{a}=\sqrt{t_{b,a}/2 n_b U_{bb}}$ with $a=1,2$. 
 In summary, we arrive
 at a purely fermionic, non-retarded description of the form:
\bea\label{Heff}
H_{{\rm eff.}} & = & \sum_{\bold{k} } \left\{ (\epsilon_{\bold{k}} -\mu_f) 
\sum_s f^\dagger_{\bold{k},s} 
f_{\bold{k},s} + \frac{U_{ff}}{V} \rho_{f,\bold{k},\uparrow}\rho_{f,-\bold{k}, \downarrow} \right.
\nonumber\\ 
&+&  \left. \frac{1}{2V} V_{{\rm ind.}, \bold{k}} \, \, \rho_{f,\bold{k},} \rho_{f,\bold{-k}} \right\}
\, ,
\eea
 This 
 is the effective Hamiltonian that we study with a numerical implementation 
 of the functional renormalization group.

\begin{figure}
\includegraphics[width=8cm]{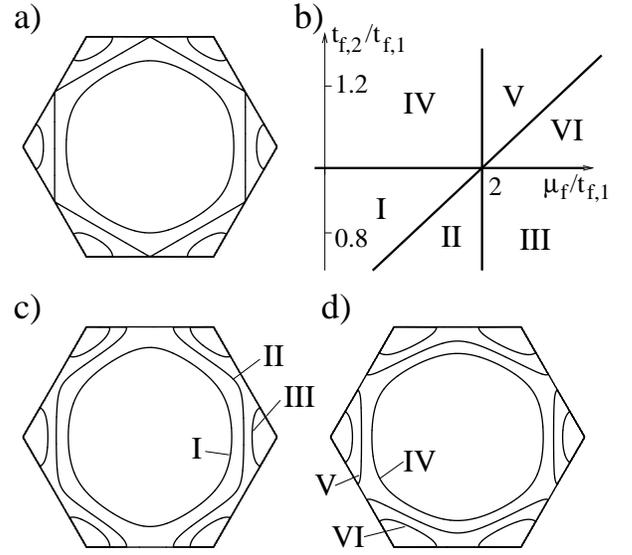}
\caption{\label{FS}
a) Fermi surfaces for an isotropic lattice, for $\mu_f<2t_{f,1/2}$,
 $\mu_f= 2t_{f,1/2}$ (hexagonal shape), 
 and $\mu_f>2t_{f,1/2}$ (six disjoint arcs).
 b) Diagram of the different types of Fermi surfaces that can be created
 on an anisotropic lattice, by varying the ratio $t_{f,2}/t_{f,1}$ 
 and $\mu_f$.
 c) Fermi surfaces for $t_{f,2}<t_{f,1}$, for $\mu_f<4t_{f,2}-2t_{f,1}$, 
 $4t_{f,2}-2t_{f,1}<\mu_f<2t_{f,1}$, and $\mu_f>2t_{f,1}$, corresponding
 to the regimes I--III, respectively.
 d)
 Fermi surfaces for $t_{f,2}>t_{f,1}$, for $\mu_f<2t_{f,1}$, 
 $2t_{f,1}<\mu_f<4t_{f,2}-2t_{f,1}$, and $\mu_f>4t_{f,2}-2t_{f,1}$, corresponding
 to the regimes IV--VI, respectively.}
\end{figure}

\section{Functional Renormalization Group}\label{FRG}
 In this section we describe 
 the  numerical implementation of functional RG \cite{shankar,zanchi} 
 that we use to study 
the effective Hamiltonian (\ref{Heff}). In this approach we integrate the
one-loop RG equations for the four-point interaction vertex
$U(\bold{k}_1,\bold{k}_2,\bold{k}_3)$, which are functions of the direction
of the momenta on the FS, 
with 
 the fourth momentum not explicitly written, given by 
${\bf k}_4 = {\bf k}_1+{\bf k}_2 -{\bf k}_3$. 
We solve the RG  equations for all the $U(\bold{k}_1,\bold{k}_2,\bold{k}_3)$ 
\cite{remark_spin}, given by:
\begin{eqnarray}
&&\partial_{\ell} U_{\ell}({\bf k_1},{\bf k_2},{\bf k_3}) =
 \nonumber\\
 &\!\!\!\!\!\!-&\!\!\!\!\int_{{\bf p},\omega}\!\!\Big\{\partial_{\ell}
  [G_{\ell}({\bf p},\omega)G_{\ell}({\bf q_1},\omega)]
 U_{\tilde{\ell}^{\prime}_{ph}}\!\!({\bf p},{\bf k_1},{\bf k_4}) U_{\tilde{\ell}^{\prime}_{ph}}\!\!({\bf p},{\bf k_3},{\bf k_2})
 \nonumber\\
 &\!\!\!\!\!\!+&\!\!\!\! 
\partial_{\ell}
 [G_{\ell}({\bf p},\omega)G_{\ell}({\bf k},-\omega)] U_{\tilde{\ell}_{pp}}\!({\bf k_1},{\bf k_2},{\bf p})
 U_{\tilde{\ell}_{pp}}\!({\bf k_3},{\bf k_4},{\bf p})
 \nonumber\\
 &\!\!\!\!\!\!+&\!\!\!\! \partial_{\ell}
  [G_{\ell}({\bf p},\omega)G_{\ell}({\bf q_2},\omega)] \left\{-\!2U_{\tilde{\ell}_{ph}}\!\!({\bf k_1},{\bf p}, {\bf k_3})U_{\tilde{\ell}_{ph}}\!\!({\bf k_4},{\bf p},{\bf k_2}) \right.
 \nonumber\\
 &\!\!\!\!\!\!+&\!\!\!\left. \!\!U_{\tilde{\ell}_{ph}}\!\!(\!{\bf p},\!{\bf k_1},\!{\bf k_3}\!) U_{\tilde{\ell}_{ph}}\!\!(\!{\bf k_4},\!{\bf p},\!{\bf k_2}\!)\! 
 +\!  U_{\tilde{\ell}_{ph}}\!\!(\!{\bf k_1},\!{\bf p},\!{\bf k_3}\!)
 U_{\tilde{\ell}_{ph}}\!\!(\!{\bf p},\!{\bf k_4},\!{\bf k_2}\!) \!\right\}\!\!\!\Big\}\!,\nonumber\\
 \end{eqnarray}
where $\ell = \ln(\Lambda_0/\Lambda)$ ($\Lambda_0 \approx 6 t_f$ is a
 high-energy cut-off), ${\bf k} \!= {\bf k}_1+{\bf k}_2-{\bf p}$, ${\bf q}_1={\bf p}+{\bf k}_1-{\bf k}_4$, ${\bf q_2}={\bf p}+{\bf k}_1-{\bf k}_3$, $\tilde{\ell}_{pp}\equiv\mbox{min}\{\ell_p,\ell_{k}\}$, $\tilde{\ell}^{\prime}_{ph}\equiv\mbox{min}\{\ell_p,\ell_{q_1}\}$,$\tilde{\ell}_{ph}\equiv\mbox{min}\{\ell_p,\ell_{q_2}\}$, $\ell_p=\ln(\Lambda_0/|\xi_{\bf p}|)$, and 
 $G_{\ell}({\bf k},\omega) = \Theta(|\xi_{{\bf k}}|-\Lambda)/(i \omega-\xi_{{\bf
     k}})$ with $\xi_{\bf k} = \epsilon_{f,{\bf k}} - \mu_f$.
The initial, i.e., bare value, of  $U(\bold{k}_1,\bold{k}_2,\bold{k}_3)$
  is given by:
\bea
U(\bold{k}_1,\bold{k}_2,\bold{k}_3) & = & U_{ff} + V_{{\rm ind.}, \bold{k}_1-\bold{k}_3} \, .
\label{Uk}
\eea
From the vertex $U(\bold{k}_1,\bold{k}_2,\bold{k}_3)$ 
we obtain the possible many-body instability channels such as CDW$_i$, spin
density wave (AF$_i$), where the index $i$ refers
 to the three nesting vectors,  
and the superconducting (BCS)  channel. 
These channels are given by:
\bea
\label{eq:cdw} V^{CDW_i} & = & 4 \, \, U_c(\bold{k}_1,\bold{k}_2, \bold{k}_1+\bold{Q}_i) \, ,
\\
\label{eq:af} V^{AF_i} & = & 4 \, \, U_\sigma(\bold{k}_1,\bold{k}_2, \bold{k}_1+\bold{Q}_i) \, ,
\\
\label{eq:bcs} V^{BCS} & = & U(\bold{k}_1,-\bold{k}_1, \bold{k}_2) \, ,
\eea
For the isotropic case, perfect nesting occurs at 3/4-filling, with three possible
nesting vectors: $\bold{Q}_1=(0,2\pi)$, $\bold{Q}_2=(\pi, \sqrt{3}\pi)$, and 
$\bold{Q}_3=(-\pi, \sqrt{3}\pi)$, leading to three different possible types
of instabilities per density wave channel.
For the anisotropic case, only $\bold{Q}_1$ can be a
 nesting vector, for the condition $\mu_f=2t_{f,1}$.

 To determine the 
 scale of the gaps, $\Delta$, associated with each of these 
order parameters, 
 we use a
 'poor man's' scaling estimate,
 specifically: $\Delta \approx \Lambda_0 e^{-\ell_c}$, where $\ell_c$
is the point at which the RG flow diverges and the instability occurs. 

The RG is implemented numerically by discretizing the FS into 
$M=24$ or $36$ patches. The CDW and AF channel is
 represented by an $(M/3)\times (M/3)$ matrix,
  the BCS channel is represented by  an 
$M \times M$ matrix, which are diagonalized at each RG step.
The dominant instability is the channel that has an eigenvalue (divided
 by the dimension of the matrix) with the
largest magnitude (for BCS one has to ensure that such eigenvalue is negative
so that the channel is attractive). Each element of the corresponding
eigenvector represents a given FS patch, and hence, the symmetry
of the dominant order parameter is reflected on the patch (i.e., angular)
dependence of each element around the FS. Using this method, we 
determine the phase diagram of the system in various limits.

\section{Isotropic lattice}\label{iso}
We first consider  spin-$1/2$ fermions on an isotropic triangular lattice, i.e. 
with  $t_{f,1}=t_{f,2}\equiv t_f$. The FS for such a lattice behaves as
follows: For small filling the FS consists of one near-circular
piece, which then approaches the shape of a hexagon as $\mu_f$ approaches
the special value $\mu_f=2t_{f}$. At this special chemical potential, which
corresponds to $3/4$-filling, the FS is nested with the three
distinct nesting vectors ${\mathbf Q}_i$. For filling fractions larger than
$3/4$ the FS breaks into six disjoined arcs. 
 Examples for these different regimes are shown in Fig. \ref{FS} a).
Without coupling to
the BEC, the fermions form an $s$-wave pairing phase for attractive interactions, and
a Fermi liquid phase for repulsive interactions \cite{KL_remark}, 
 except for the specific case 
$\mu_f = 2 t_{f}$,  where the system shows AF order for repulsive interactions.
A schematic picture of this order is shown in Fig. \ref{phases} b) 
for the nesting vector $\bold{Q}_1$. 

We found a similar behavior for an isotropic square lattice
 in [\onlinecite{bfm_square}], $s$-wave pairing for attractive interaction,
 and Fermi liquid behavior for repulsive interaction, except 
 at half-filling, for which we find AF order.
 An interesting difference for the triangular lattice
 is the three-fold degeneracy of the AF phase, an indication
 of frustration.

\begin{figure}
\includegraphics[width=8cm]{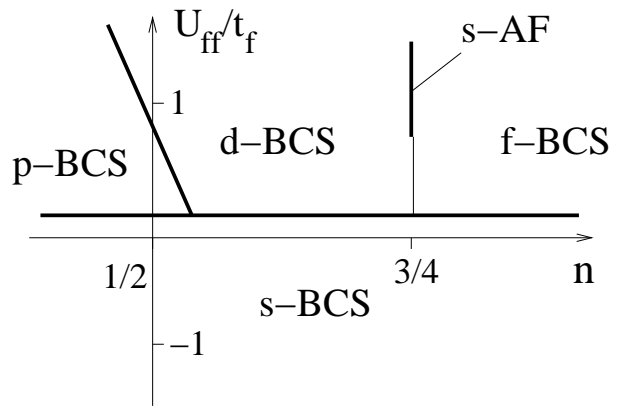}
\caption{\label{PD}
Phase diagram
 of a Bose-Fermi mixture on a 2D isotropic triangular lattice.
 The vertical axis corresponds to the
  interaction strength, $U_{ff}/t_f$, whereas 
 the horizontal axis  corresponds to the
 filling fraction of the fermions per site, $n$. 
 The other parameters are given by $\tilde{V}/t_f = 3$, and $\xi_{a} = 1$
with $a=1,2$.}
\end{figure}

When one turns on the coupling to the
BEC, the isotropic system shows a phase diagram of the type shown in Fig. \ref{PD}.
The $s$-wave pairing  phase slightly extends into the regime of positive $U_{ff}$,
because of the induced attractive interaction mediated by the bosonic fluctuations.
The regime that showed Fermi liquid behavior in the absence of the induced
interaction now shows a rich competition of various types of pairing.
In the regime where the density is below half-filling,
when the FS is approximately circular, the system shows $p$-wave pairing.
For fillings larger than $3/4$, when the FS consists of six disjoined parts,  
the fermions Cooper pair in a superconducting state with $f$ symmetry.
As shown in Fig. \ref{phases} e), the FS in this regime can also
be interpreted as two distinct near-circular FSs of holes. 
In this interpretation each of these two fermionic systems is in an $s$-wave
pairing phase, but the relative phase between the two order parameters is $\pi$.
At $3/4$-filling and large values of $U_{ff}$, the system still shows AF
order. However, for smaller values of $U_{ff}$, and also for smaller
fillings, two phases with degenerate extended $d$ symmetry
develop. These superconducting orders have a sizeable $g$-wave component and
are approximately given by:
\begin{eqnarray}
\psi_{D_1} & = & \sin 2\theta + 0.5 \sin 4 \theta\\
\psi_{D_2} & = & \cos 2\theta - 0.5 \cos 4 \theta
\label{d1d2}
\end{eqnarray}
These order parameters are shown in Fig. \ref{phases} c) and d).
The shapes of the order parameters are energetically 
advantageous because, on the one hand, the order parameter maxima are located
at points at which the system has a high density of states 
(the 'corners' of the FS). Hence, when the superconducting gap opens, 
there is a large gain of condensation energy coming from these regions on
the FS.  On the other hand, the $d$-wave state
has lower kinetic energy than the $f$-wave, and hence is selected.

The phase diagram Fig. \ref{PD}
 has a number of 
similarities to the phase diagram for a BFM on a square lattice,
 such as the $s$- and the $p$-wave pairing phase, 
  and the existence of AF order 
 for a nested Fermi surface for large $U_{ff}$.
 However,
 the competition of pairing orders
 for positive $U_{ff}$ and
 intermediate  and large filling
 is much richer, due to the 
 more complex shape of the 
 Fermi surface. 

The energy gaps associated with these order parameters
 can be determined as we did in [\onlinecite{bfm_square}], by using
 a 'poor man's' scaling argument.
 We find for the $s$-wave pairing and the AF order, that they are
 around $0.1 T_F$, 
where $T_F$ is the Fermi temperature of the system.
 For most of the exotic phases, we energy gaps of the order of
 $0.01 - 0.001 T_F$.

\section{Anisotropic Lattice}\label{aniso}
We now consider a BFM with spin-$1/2$ fermions on an anisotropic 
 triangular lattice, 
i.e. with unequal hopping amplitudes, $t_{f (b),1} \neq t_{f (b),2}$.
The shape of the FS behaves as follows:  For $t_{f,2}>t_{f,1}$, 
as one increases the chemical potential, the FS first breaks 
into four arcs at $\mu_{f}=2 t_{f,1}$, and then breaks into six arcs
at $\mu_{f} = 4t_{f,2}-2 t_{f,1}$,
 corresponding to the regimes IV--VI, in Fig. \ref{FS} b) and d). 
For $t_{f,2}<t_{f,1}$ the FS 
first breaks into two arcs at $\mu_{f} = 4t_{f,2}-2 t_{f,1}$, and then
breaks into six arcs at $\mu_f=2t_{f,1}$
 coresponding to the regimes I--III, in Fig. \ref{FS} b) and c). 
At the special chemical potential
$\mu_f=2t_{f,1}$ the FS is still nested,  but there is only one nesting vector 
along the direction of the bonds with hopping amplitude $t_{f,1}$.
In the absence of the coupling to the BEC the phase diagram has a similar
structure as for the isotropic case: $s$-wave pairing for attractive
interaction, Fermi liquid behavior for repulsive interaction, with the exception 
of the nested FS at $\mu_f=2t_{f,1}$ where one finds AF order (notice that
in this case the filling is not $3/4$).
  
\begin{figure}
\includegraphics[width=8cm]{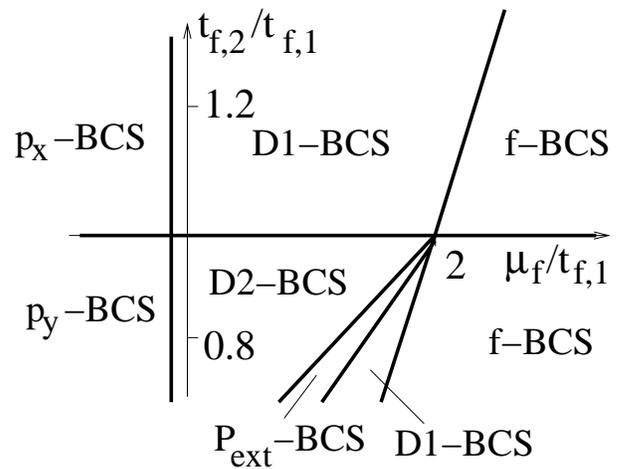}
\caption{\label{PDaniso}
Phase diagram of  a Bose-Fermi mixture on an anisotropic triangular lattice.
 The vertical axis corresponds to
 the ratio $t_{f,2}/t_{f,1}$, the horizontal axis
 corresponds to the chemical potential $\mu_f$.  
 The other parameters are given by $\tilde{V}/t_f = 3$, 
 $U_{ff}/t_{f,1}=2$, and  $\xi_{a} = 1$ with $a=1,2$).}
\end{figure}

When the coupling to the bosons is turned on, one generates an even more
complicated competition of pairing phases for repulsive $U_{ff}$
in the vicinity of the point $\mu_f = 2t_{f,1}$, as is shown
in Fig. \ref{PDaniso}. Generally, for unequal hopping the degeneracy between 
$D_1$ and $D_2$ in (\ref{d1d2}), as well as $p_x$ and $p_y$ is lifted: In the
regime with $t_{f,2}>t_{f,1}$ ($t_{f,2}<t_{f,1}$), $D_1$ ($D_2$) and $p_x$
($p_y$) dominate. For $t_{f,2}>t_{f,1}$, in the intermediate regime, in which the
 FS consists of four disjoined arcs, 
 corresponding to the regime V in Fig. \ref{FS}, 
the type of ordering changes from $D_1$ to $f$.  
 For $t_{f,2}<t_{f,1}$, the type of pairing also eventually becomes $f$-wave,
 but first develops two other types of pairing, in
 the regime II in Fig. \ref{FS}. 
 Firstly, one finds an unusual
 extended $p$-wave symmetry, which is schematically shown in Fig. \ref{phases} f).
 Its wavefunction is of the form:
\bea
\psi_{P_{ext}} 
& = &\left\{
\begin{array}{cr}
 \sin^2 \theta  & -\pi/2 < \theta <\pi/2\\
 -\sin^2 \theta &  \pi/2 < \theta <3\pi/2
\end{array}
\right.  
\eea
The second type of pairing that appears before the system develops $f$-wave 
pairing is $D_1$. These unusual pairing states are energetically favorable
because of the 
 anisotropic shape of the FS. For the regime in which the FS has just 
barely broken up into two arcs, the order parameter assumes $p$-wave symmetry
and the maxima are located along the $y$-axis, where the density of states is highest.
As the region of open FS widens (see Fig. \ref{phases} f)), 
this pairing becomes energetically unfavorable, and
the system develops $D_1$-pairing, so that the maxima of the order parameter 
can again be located near the point of highest density of states. 
 The energy gaps associated 
 with these order parameters are
 of the same order of magnitude as  
 for 
 the isotropic lattice.

\section{Spinless Fermions}\label{nospin}  
Finally, we consider a BFM with spinless fermions on an isotropic
lattice. Due to the absence of $s$-wave scattering between fermions
of the same spin state, there is no direct interaction, that is, $U_{ff}=0$.
Hence, in the absence of bosons, the spinless gas is non-interacting. 
The boson fluctuations, however, mediate an induced interaction between the fermions. 
Due to the anti-symmetry of the Cooper pair wavefunction, pairing occurs 
in an odd angular momentum channel.  We find a competition between $p$ and 
$f$-wave pairing symmetry. For small to intermediate filling ($n<0.65$), 
$p$-wave pairing dominates. For larger fillings, for which the FS first approaches
the shape of a hexagon and then breaks up into six arcs, the system
shows $f$-wave pairing. 
     Since  these larger fillings of fermions are typically realized in the center of an atomic trap,
 this result would suggest a comparatively easy way to create an exotic pairing state experimentally.  

In contrast to this,
 a spinless BFM on a square lattice
 only shows $p$-wave pairing, since for
 the quadrangular shape of its
 FS, channels of higher angular momentum 
 are of no advantage energetically. 

\section{Conclusion}\label{conclude}
%
We have used a functional RG approach to study BFM 
of ultra-cold atoms, both with spin-$1/2$ and with spinless fermions, 
in a 2D optical lattice with triangular geometry. 
 We found a number of 
competing types of order:
 For an isotropic lattice, we found 
  pairing states with $s$, $p$-,  
 extended $d$-, and $f$-wave symmetry, as well as AF order. 
 For an anisotropic lattice, we additionally found
  extended $p$-wave pairing, and that the degeneracy
 between the two types of extended $d$-wave pairing and
 the two types of $p$-wave pairing has been lifted.
 For a BFM with spinless fermions, we found
 a competition between $p$- and $f$-wave pairing.
 From the RG flow 
 we also identified the magnitude of 
 the superconducting gaps, which
 turn out to be a fraction of the Fermi temperature, of the order of
 $0.01 T_F$--$0.001T_F$. 
Nevertheless, it is possible to increase these gaps substantially if the
coupling to the BEC is in the strong-coupling regime. In this case,
retardation effects can enhance the pairing mechanisms substantially
requiring an Eliashberg theory instead of BCS. 
To take into account retardation effects in this problem, the frequency
dependence of $U$ should also be kept \cite{tsai}. The functional RG method 
with frequency dependence can be used in this regime.

An important question in the context of
 ultracold atoms concerns the detectability of these phases. 
The exotic superconducting and AF states discussed in this work exhibit distinct
experimental signatures. The AF state can be studied via 
time-of-flight images and Bragg scattering \cite{Bragg}, 
and the different superconducting phases can be measured through 
noise correlations \cite{noise}. Finally, laser stirring experiments 
\cite{stirring} can be used to establish the phase boundary between two
phases.

In conclusion, we presented a systematic study of the weak-coupling limit
 of BFMs of ultracold atoms in triangular optical lattices.
 For the
  various regimes, 
 we found a number of exotic phases, and discussed their  
 properties  and their relevance for ultra-cold atom experiments. 
Given the generic structure of the 
 Hamiltonian that we considered, the new types of phases, such as the extended $d$-wave pairing phase and the extended $p$-wave phases, are also of interest
 for the study of solid-state systems with triangular symmetry.

We thank A.~Polkovnikov for illuminating conversations. 
A.~H.~C.~N. was supported by the NSF grant DMR-0343790.



\begin{thebibliography}{99}
\bibitem{remark_spin}
$U(\bold{k}_1,\bold{k}_2,\bold{k}_3)$ denotes processes
involving electrons with opposite spins. Processes involving parallel spins 
can be obtained from these \cite{zanchi}.
\bibitem{KL_remark}
We ignore high angular momentum 
pairing phases predicted by the Kohn-Luttinger theorem which would 
occur at energy scales much lower than the experimentally accessible
 regime.
\end{thebibliography}
\end{document}